\documentclass[12pt,a4paper, superscriptaddress]{article}
\usepackage{epsfig,amsmath,amssymb}
\usepackage[colorlinks=true,linktocpage=true,linkcolor=blue,citecolor=blue]{hyperref}
\usepackage[colorlinks=true,linktocpage=true,linkcolor=blue,citecolor=blue]{hyperref}
\usepackage{cite}
\usepackage{authblk}
\usepackage{multirow}
\usepackage{graphicx}
\usepackage{dcolumn}
\usepackage{bm}
\usepackage{subfig}
\usepackage{float}
\usepackage{booktabs}
\usepackage{tabularx}
\usepackage{appendix}
\usepackage{hhline}
\usepackage{ulem}
\usepackage{color}
\usepackage{slashed}
\title{Medium-Induced Quarkonium Dissociation at Finite Chemical Potential and Weak Magnetic Field}
\date{}
\author[1]{Indrani Nilima\thanks{nilima.ism@gmail.com}}
\author[2]{Mujeeb Hasan\thanks{mhasan@lords.ac.in}}
\author[3]{Mohammad Yousuf Jamal\thanks{yousufjml5@gmail.com}}
\author[4]{Salman Ahamad Khan\thanks{salmankhan.dx786@gmail.com}}
\author[5]{B. K. Singh\thanks{bksingh@bhu.ac.in}}

\affil[1]{Discipline of Natural Sciences, PDPM IIITDM, Jabalpur 482005, India}
\affil[2]{Department of Physics, Lords Institute of Engineering and Technology, Hyderabad 500091, Telangana, India}
\affil[3]{Key Laboratory of Quark and Lepton Physics (MOE) \& Institute of Particle Physics, Central China Normal University, Wuhan 430079, China}
\affil[4]{Department of Physics, Integral University, Lucknow 226026, India}
\affil[1,5]{Department of Physics, Institute of Science, Banaras Hindu University (BHU), Varanasi 221005, India}

\begin{document}

\maketitle

\begin{abstract}
We investigate the in-medium modification and dissociation of heavy quarkonium in a hot QCD medium at finite quark chemical potential and in the weak magnetic-field regime. Starting from the one-loop resummed gluon propagator in the imaginary-time formalism, and incorporating non-perturbative effects through a phenomenological correction to the HTL description, we compute the real and imaginary parts of the dielectric permittivity. This, in turn, leads to a complex heavy-quark potential: the real part is used to determine binding energies by solving the nonrelativistic Schr\"odinger equation, while the imaginary part generates thermal decay widths, dominated by Landau damping. Within the explored parameter range, temperature has the greatest control over Debye screening, potential modification, and quarkonium stability, whereas finite density and weak magnetic fields introduce comparatively smaller quantitative changes. As the temperature increases, binding energies decrease and thermal widths grow, giving rise to the expected hierarchy between ground and excited states and a sequential suppression pattern in the dissociation temperatures. Overall, our results indicate that while finite chemical potential and weak magnetic fields can shift quarkonium properties in a measurable way, thermal effects remain the primary driver of dissociation, with direct relevance for heavy-ion collision phenomenology.
\end{abstract}
\maketitle
\section{Introduction}
In extreme conditions of temperature and/or density, hadronic matter is expected to deconfine into a plasma of quarks and gluons, commonly referred to as the quark--gluon plasma (QGP). Studying this phase is important for understanding the early Universe as well as the physics of compact stars. In the laboratory, QGP is produced in ultra-relativistic heavy-ion collisions at RHIC (BNL)~\cite{Arsene:NPA'2005,Adcox:NPA'2005} and the LHC (CERN)~\cite{Carminati:JPG'2004,Alessandro}. Among the proposed signatures of this short-lived, strongly interacting medium, heavy quarkonia (bound states of \(c\bar{c}\) and \(b\bar{b}\)) are particularly valuable because the inter-quark potential is Debye screened in the medium, which can lead to quarkonium suppression~\cite{Matsui:PLB178'1986}.

Over the last few decades, heavy quark bound states have been investigated within effective field theories that exploit the hierarchy of energy scales in a \(Q\bar{Q}\) system. Non-relativistic QCD (NRQCD) is obtained by integrating out the heavy-quark mass scale~\cite{Bodwin:PRD51'1995}, while potential NRQCD is formulated by integrating out the momentum-exchange scale~\cite{Brambilla:NPB566'2000}. Quarkonium spectroscopy has also been studied from first principles using lattice QCD~\cite{Braguta:PRD89'2014}. Additionally, phenomenological potential models have been widely employed to investigate the in-medium properties of quarkonium. A key result is that the in-medium \(Q\bar{Q}\) potential becomes complex, with both real and imaginary parts~\cite{Laine:JHEP, Nilima:2024qzx, Nilima:2024nvd, Singh:2023zxu, Sebastian:2022sga, Jamal:2018mog, Agotiya:2016bqr}. The real part is screened due to Debye screening, whereas the imaginary part gives rise to a finite thermal width. Medium modifications of the \(Q\bar{Q}\) potential have been studied by including both the perturbative Coulomb component~\cite{Dumitru:PRD79'2009} and the non-perturbative string component~\cite{Thakur89,Thakur:PRD88'2013} of the Cornell potential. The imaginary part of the potential has also been computed using the AdS/CFT correspondence~\cite{Patra:PRD92'2015,Patra:PRD91'2015}. The authors of~\cite{Rothkopf:PRL108'2012} extracted both the real and imaginary parts of the potential from thermal Wilson-loop spectral functions using lattice QCD. In a related study~\cite{Lafferty:PRD101'2020}, the generalized Gauss law was employed, with divergences regulated through the Debye mass. More recently, non-perturbative effects in the potential have been incorporated via a non-perturbative propagator that accounts for low-frequency modes subsumed in dimension-two gluon condensates~\cite{Guo:PRD100'2019}.

A strong transient magnetic field is produced in heavy-ion collisions due to the relative motion of spectator charges. Its magnitude is estimated to be of order \(m_{\pi}^{2}\) (\(\sim 10^{18}\) Gauss) at RHIC and of order \(10\,m_{\pi}^{2}\) at the LHC~\cite{Skokov:IJMPA24'2009,Voronyuk:PRC83'2011}. In recent years, the implications of a magnetic field for the properties of QCD matter have been explored extensively~\cite{Braguta:PRD89'2014,Fukushima:PRD78'2008,Nilima:2022tmz,Kharzeev:PRL106'2011,Gusynin:PRL73'1994}. Although the magnetic field decays rapidly in vacuum, its lifetime can be significantly extended in a conducting medium due to finite electrical conductivity~\cite{Tuchin:AHEP2013'2013,Mclerran:NPA929'2014,Rath:PRD100'2019}. Since the characteristic quarkonium formation time scale (\(\sim 1/2m_{Q}\), where \(m_{Q}\) is the charm or bottom quark mass) is comparable to the time scale over which the magnetic field can remain sizable, it is important to assess the impact of the background magnetic field on \(Q\bar{Q}\) bound states. The effect of a magnetic field on QCD bound states has been studied with a harmonic interaction and a spin--spin interaction term, in addition to the usual Cornell potential, in\cite{Alford:PRD88'2013,Bonati:PRD92'2015}. The in-medium properties of heavy quark bound states have also been investigated in strong magnetic field in~\cite{bonati:PRD94'2016,bonati:PRD95'2017,hasan:EPJC77'2017,Mujeeb:NPA995'2020,Khan:NPA1034'2023} and weak magnetic field~\cite{Mujeeb:PRD102'2020}. The complex heavy-quark potential in a magnetic background has been studied using the generalized Gauss law~\cite{Singh:PRD97'2018}. Along with magnetic effects, a sizeable quark chemical potential (up to \(\mu \sim 100\) MeV) can be produced near the critical temperature\cite{Braun:JPG28'2002,Cleymans:JPG35'2008,Andronic:NPA837:2010}, and in the presence of a strong magnetic field it can reach \(\sim 200\) MeV~\cite{Fukushima:PRL117'2016}. The study of strongly interacting matter at finite baryon density is essential for the physics of the early Universe and compact stars. While the QCD phase diagram at vanishing chemical potential has been extensively explored using lattice and analytical methods, it remains considerably less constrained at finite chemical potential due to the sign problem in lattice calculations. Such baryon-rich matter is expected to be created in forthcoming experiments at FAIR. Since in-medium \(Q\bar{Q}\) properties provide important information about the surrounding medium, it is therefore timely to examine the combined influence of finite chemical potential and magnetic field on these bound states.

In the present work, we study the effects of finite quark chemical potential on the properties of \(Q\bar{Q}\) bound states immersed in a weakly magnetized hot QCD medium. We first compute the one-loop resummed gluon propagator using the imaginary-time formalism in the hard thermal loop approximation. To incorporate non-perturbative contributions, we add a phenomenological term to the resummed HTL propagator that captures low-energy effects. The static limit of the \(00\)-component of the resummed gluon propagator provides the dielectric permittivity of the medium, which is then used to determine the medium modification of the heavy-quark potential. The binding energies of charmonium and bottomonium states are obtained by solving the Schr\"odinger equation with the real part of the potential, while the imaginary part is used to compute the thermal width. The dissociation temperature is inferred from the interplay between the binding energy and the thermal width.

The paper is organized as follows. In section~\ref{potential}, we discuss the medium modification of the complex heavy-quark potential in the presence of a weak magnetic field, chemical potential, and temperature. In subsection~\ref{complexpermittivity}, we evaluate the real and imaginary parts of the dielectric permittivity, which lead to the real and imaginary parts of the heavy-quark potential in subsection~\ref{complexpotential}. In section~\ref{properties}, we calculate the binding energy, thermal width, and dissociation of quarkonia. In section~\ref{result}, we present and discuss our results. Finally, we summarize our findings in section~\ref{conclusion}.

\section{Complex Heavy Quarkonium Potential}
\label{potential}

In this section, we discuss the in-medium modification of the interaction potential between a heavy quark \(Q\) and its antiquark \(\bar{Q}\) in a hot QCD medium at finite chemical potential and in the presence of a weak magnetic field. Owing to the large heavy-quark mass \(m_Q\), the scale hierarchy
\[
m_Q \gg T \gg \Lambda_{\mathrm{QCD}}, \qquad
m_Q \gg \sqrt{eB}, \qquad
m_Q \gg \mu
\]
is well satisfied in the regime of interest. This separation of scales justifies a potential description of the \(Q\bar{Q}\) pair in the static limit, where medium effects can be encoded through an effective (complex) potential.

We incorporate medium modifications through a dielectric function \(\epsilon(q)\) as
\begin{equation}
V(r;T,\mu,B)=\int\frac{d^3q}{(2\pi)^{3/2}}
\left(e^{i\mathbf{q}\cdot \mathbf{r}}-1\right)\frac{V(q)}{\epsilon(q)},
\label{pot_defn}
\end{equation}
where the \(r\)-independent subtraction term renormalizes the heavy-quark free energy (the free energy at infinite separation). The Fourier transform of the perturbative Coulomb part of the Cornell potential \(V(r)=-\frac{4\alpha_s}{3r}\) reads ~\cite{Nilima:2022tmz}
\begin{equation}
V(q)=-\frac{4}{3}\sqrt{\frac{2}{\pi}}\,\frac{\alpha_s}{q^2}.
\label{ft_pot}
\end{equation}
We emphasize that we start from the perturbative part of the vacuum potential only. The reason is that the same screening scale cannot be applied in a straightforward way to both the Coulomb and string terms due to the intrinsically non-perturbative nature of the string contribution. In what follows, the non-perturbative part is incorporated through a dimension-two gluon condensate prescription, as described in Ref.~\cite{Guo:PRD100'2019}.

\subsection{Complex permittivity in a hot QCD medium with weak magnetic field}
\label{complexpermittivity}

The (complex) dielectric permittivity is obtained from the static limit of the \(00\)-component of the resummed gluon propagator via linear response theory:
\begin{equation}
\frac{1}{\epsilon(q)}=-\lim_{q_0\rightarrow 0} q^{2}\, D^{00}(q_0,q).
\label{dielectric}
\end{equation}
The general form of the resummed gluon propagator in Landau gauge can be written as~\cite{karmakar:EPJC79'2019}
\begin{eqnarray}
D^{\mu\nu}(Q)=\frac{(Q^2-d)B^{\mu\nu}}{(Q^2-b)(Q^2-d)-a^2}
+\frac{R^{\mu\nu}}{Q^2-c}
+\frac{(Q^2-b)M^{\mu\nu}}{(Q^2-b)(Q^2-d)-a^2}
+\frac{aN^{\mu\nu}}{(Q^2-b)(Q^2-d)-a^2},
\end{eqnarray}
where \(B^{\mu\nu}\), \(R^{\mu\nu}\), \(M^{\mu\nu}\), and \(N^{\mu\nu}\) are tensor structures and \(a\), \(b\), and \(d\) are form factors~\cite{karmakar:EPJC79'2019}. Since \(R^{00}=M^{00}=N^{00}=0\), the \(00\)-component reduces to
\begin{eqnarray}
D^{00}(Q)=\frac{(Q^2-d)\,\bar{u}^2}{(Q^2-b)(Q^2-d)-a^2}.
\label{propagator_00}
\end{eqnarray}

The form factor \(a(Q)\) can be decomposed as
\begin{eqnarray}
a(Q)=a_0(Q)+a_2(Q),
\end{eqnarray}
where \(a_0\sim O(q_fB)^0\) and \(a_2\sim O(q_fB)^2\). The zero-field contribution vanishes, \(a_0=0\), and \(a_2\) enters the propagator denominator through \(a^2\), i.e. at \(O(q_fB)^4\). Since we retain terms up to \(O(q_fB)^2\) in the weak-field expansion, the contribution of \(a\) can be neglected consistently. Therefore, up to \(O(q_fB)^2\),
\begin{eqnarray}
D^{00}(Q)=\frac{\bar{u}^2}{(Q^2-b)}.
\label{propagator_final}
\end{eqnarray}

\subsubsection{Real and imaginary parts of the resummed propagator}

Using Eq.~\eqref{propagator_final}, the real part in the static limit becomes
\begin{eqnarray}
{\rm \Re}\,D^{00}(q_0=0,q)=-\frac{1}{q^2+m_D^2},
\label{real_resummed}
\end{eqnarray}
where \(m_D^2={\rm \Re}\,b(q_0=0,q)\) defines the Debye screening mass in the present medium.

For the imaginary part, one may express \({\rm \Im}\,D^{00}\) in terms of the real and imaginary parts of the form factor \(b\), using the standard relation~\cite{Weldon:PRD42'1990}
\begin{eqnarray}
{\rm \Im}\,D^{00}(q_0,q)=\frac{2T}{q_0}\,
\frac{{\rm \Im}\,b(q_0,q)}
{(Q^2-{\rm \Re}\,b(q_0,q))^2+({\rm \Im}\,b(q_0,q))^2}.
\end{eqnarray}
Rewriting it as
\begin{eqnarray}
{\rm \Im}\,D^{00}(q_0,q)=2T\,
\frac{\left[\frac{{\rm \Im}\,b(q_0,q)}{q_0}\right]}
{(Q^2-{\rm \Re}\,b(q_0,q))^2+\left(q_0\left[\frac{{\rm \Im}\,b(q_0,q)}{q_0}\right]\right)^2},
\end{eqnarray}
and taking the static limit yields
\begin{eqnarray}
{\rm \Im}\,D^{00}(q_0=0,q)=2T\,
\frac{\left[\frac{{\rm \Im}\,b(q_0,q)}{q_0}\right]_{q_0=0}}
{(q^2+m_D^2)^2}.
\label{resummed}
\end{eqnarray}

\subsubsection{Debye mass and running coupling}
\begin{figure}
\begin{center}
\includegraphics[width=4.6cm,height=4.8cm]{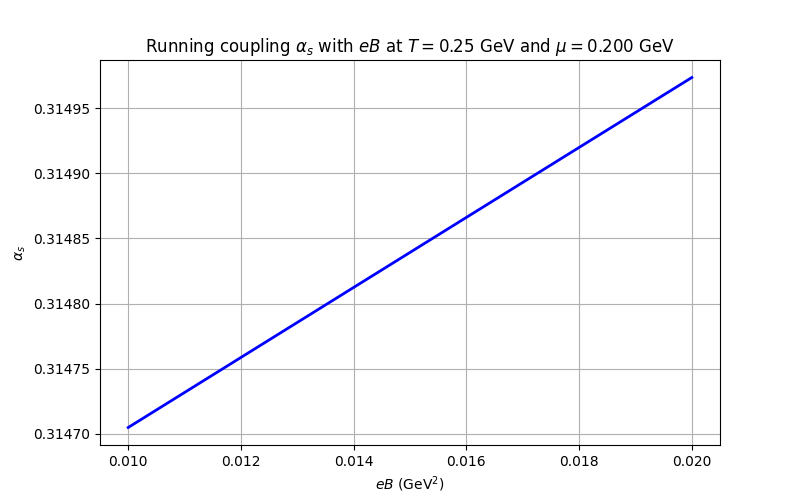}
\hspace{5mm}
\includegraphics[width=4.6cm,height=4.8cm]{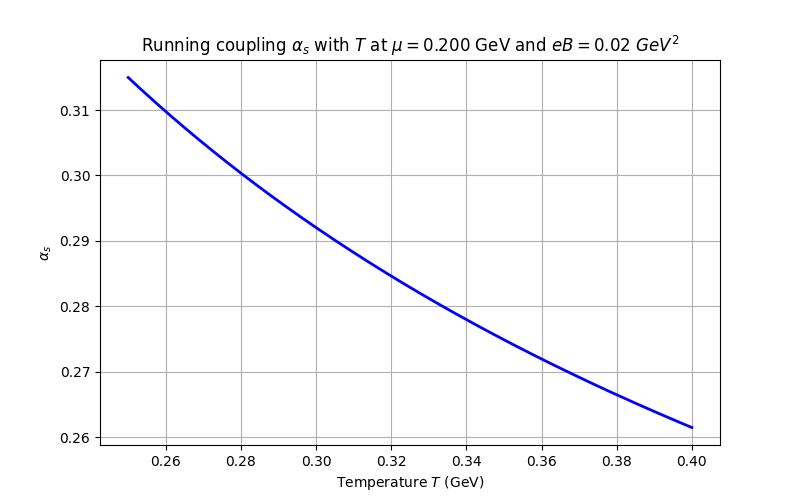}\\
\caption{Variation of coupling with magnetic field (left panel) and with temperature (right panel).}
\label{alphat}
\end{center}
\end{figure}

In a weakly magnetized medium at finite chemical potential, the Debye mass is taken as~\cite{Bandyopadhya:PRD100'2019}
\begin{eqnarray}
m_D^2=\frac{g^2T^2}{3}\left\lbrace\left(N_c+\frac{N_f}{2}\right)+6N_f\hat{\mu}^2\right\rbrace+
\sum_f\frac{g^2}{12\pi^2 T^2}(q_fB)^2
\sum_{l=1}^{\infty}(-1)^{l+1}l^2\cosh(2l\pi\hat{\mu})K_0\!\left(\frac{m_fl}{T}\right),
\label{debyemass}
\end{eqnarray}
where \(\hat{\mu}=\frac{\mu}{2\pi T}\), \(q_f\) and \(m_f\) denote the electric charge and mass of quark flavor \(f\), and \(K_0\) is the modified Bessel function. The strong coupling is taken to run with \((T,\mu,eB)\) as~\cite{ayala:PRD98'2018}
\begin{eqnarray}
\alpha_s(\Lambda^2,eB)=\frac{g^2}{4\pi}=\frac{\alpha_s(\Lambda^2)}{1+
b_1\alpha_s(\Lambda^2)\ln\left(\frac{\Lambda^2}{\Lambda^2+eB}\right)},
\end{eqnarray}
with
\begin{eqnarray}
\alpha_s(\Lambda^2)=\frac{1}{b_1\ln\left(\frac{\Lambda^2}{\Lambda_{\overline{MS}}^2}\right)},
\end{eqnarray}
where \(\Lambda=2\pi \sqrt{\left(T^2+\frac{\mu^2}{\pi^2}\right)}\), \(b_1=\frac{11N_c-2N_f}{12\pi}\), and \(\Lambda_{\overline{MS}}=0.176~{\rm GeV}\). In Fig.~\ref{alphat}, we show the variation of \(\alpha_s\) with magnetic field and temperature.

Using the corresponding expression for \(\left[\frac{{\rm \Im}\,b}{q_0}\right]_{q_0=0}\) together with Eq.~\eqref{resummed}, the imaginary part of the \(00\)-component can be written in the compact form
\begin{eqnarray}
{\rm \Im}\,D^{00}(q_0=0,q)=\frac{\pi T\,M^2_{(T,\mu,B)}}{q\,(q^2+m_D^2)^2},
\label{img_resummed}
\end{eqnarray}
where \(M^2_{(T,\mu,B)}\) is defined as
\begin{eqnarray}
M_{(T,\mu,B)}^2&=&\frac{g^2T^2}{3}\left\lbrace\left(N_c+\frac{N_f}{2}\right)+6N_f\hat{\mu}^2\right\rbrace
+\left[\sum_f\frac{g^2(q_fB)^2}{8\pi^2 T^2}
\sum_{l=1}^\infty(-1)^{l+1}l^2K_0\!\left(\frac{m_f l}{T}\right)
\right. \nonumber\\ &&-\left.
\sum_f\frac{g^2(q_fB)^2}{48\pi^2 T^2}
\sum_{l=1}^\infty(-1)^{l+1}l^2K_2\!\left(\frac{m_f l}{T}\right)
+\sum_f\frac{g^2(q_fB)^2}{384\pi^2}\frac{(8T-7\pi m_f)}{m_f^2 T}
\right].
\label{Mt}
\end{eqnarray}

\subsubsection{Non-perturbative correction and \(\epsilon(q)\)}

Following Ref.~\cite{Guo:PRD100'2019}, we incorporate non-perturbative effects through a dimension-two gluon condensate contribution in the propagator. The non-perturbative (NP) terms in the static limit are taken as
\begin{eqnarray}
{\rm \Re}\,D^{00}_{\rm NP}(q_0=0,q)=-\frac{m_G^2}{(q^2+m_D^2)^2},\\
{\rm \Im}\,D^{00}_{\rm NP}(q_0=0,q)=\frac{2\pi T\,M^2_{(T,\mu,B)}\,m_G^2}{q\,(q^2+m_D^2)^3},
\end{eqnarray}
where \(m_G^2\) is a dimensionful constant. It can be related to the string tension through \(\sigma=\alpha\,m_G^2/2\).

Including both HTL and NP contributions, the real and imaginary parts of \(D^{00}\) are
\begin{eqnarray}
{\rm \Re}\,D^{00}(q_0=0,q)&=&-\frac{1}{q^2+m_D^2}-\frac{m_G^2}{(q^2+m_D^2)^2},
\label{real_propagator}\\
{\rm \Im}\,D^{00}(q_0=0,q)&=&\frac{\pi T\,M^2_{(T,\mu,B)}}{q\,(q^2+m_D^2)^2}
+\frac{2\pi T\,M^2_{(T,\mu,B)}\,m_G^2}{q\,(q^2+m_D^2)^3}.
\label{imaginary_propagator}
\end{eqnarray}
\begin{figure}
\begin{center}
\includegraphics[width=4.6cm,height=4.8cm]{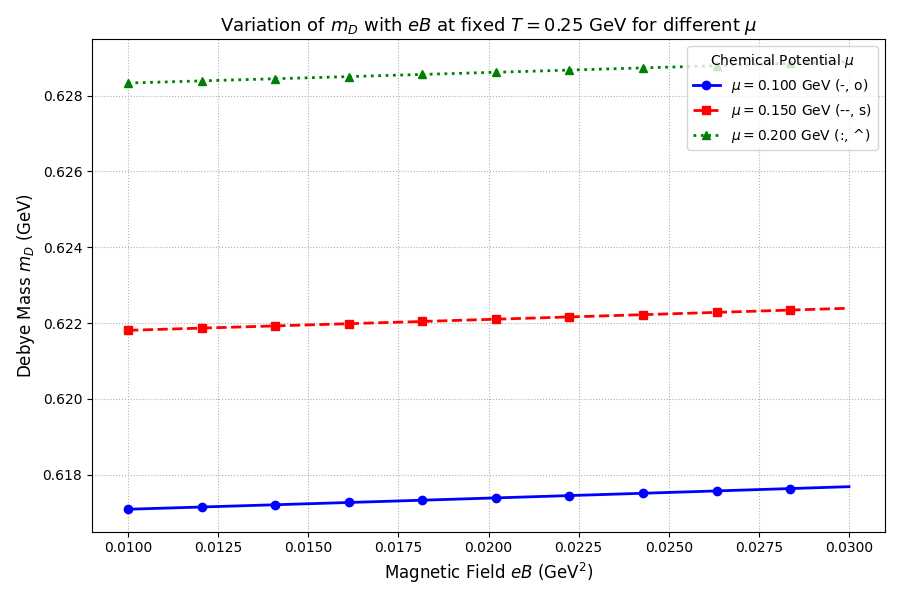}
\includegraphics[width=4.6cm,height=4.8cm]{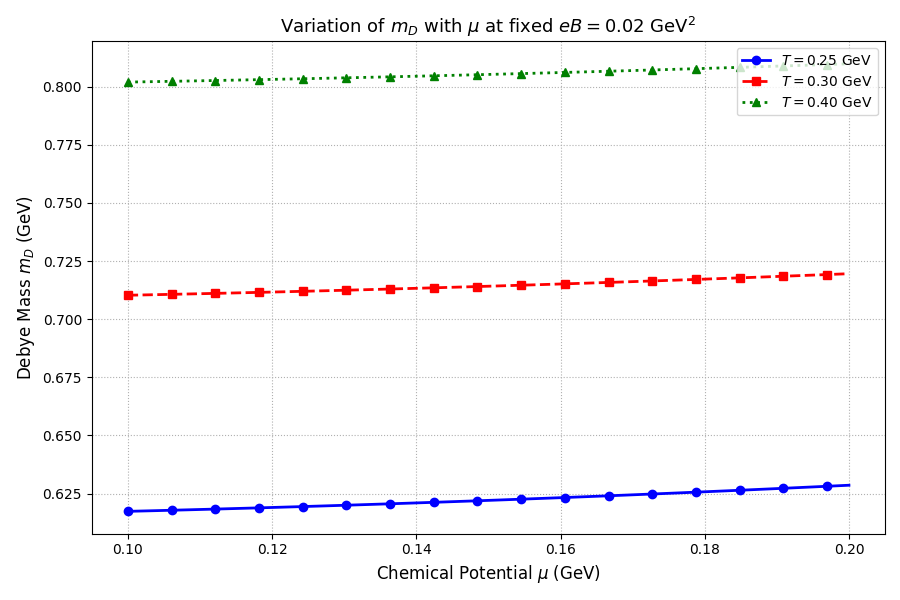}
\includegraphics[width=4.6cm,height=4.8cm]{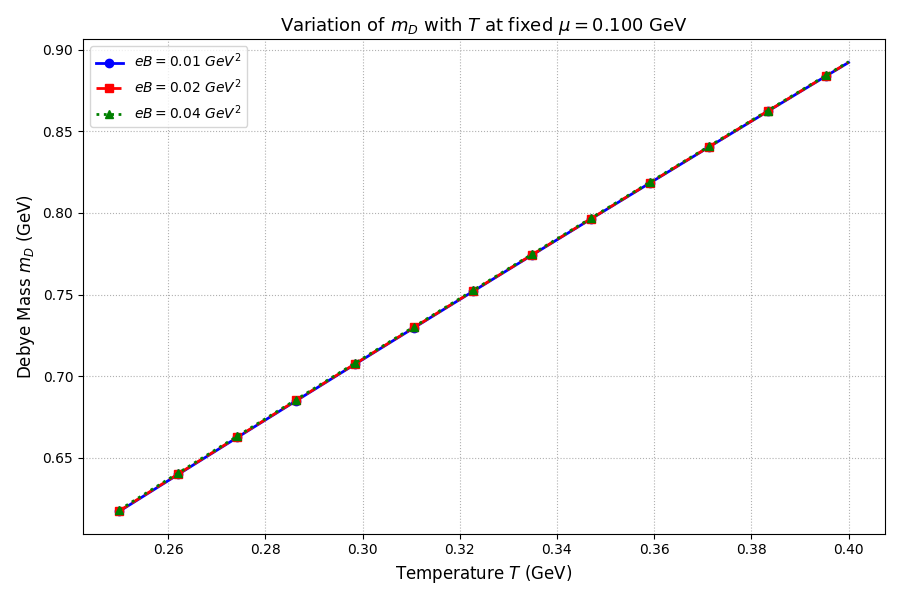}
\caption{Variation of Debye mass with magnetic field (left panel), with chemical potential (middle panel), and with temperature (right panel).}
\label{debye}
\end{center}
\end{figure}

From Eq.~\eqref{dielectric}, it is the real and imaginary parts of \(1/\epsilon(q)\) that enter the potential:
\begin{eqnarray}
{\rm \Re}\left[\frac{1}{\epsilon(q)}\right]&=&\frac{q^2}{q^2+m_D^2}+\frac{q^2 m_G^2}{(q^2+m_D^2)^2},
\label{real_dielectric}\\
{\rm \Im}\left[\frac{1}{\epsilon(q)}\right]&=&-\frac{q\pi T\,M^2_{(T,\mu,B)}}{(q^2+m_D^2)^2}
-\frac{2q\pi T\,M^2_{(T,\mu,B)}\,m_G^2}{(q^2+m_D^2)^3}.
\label{imaginary_dielectric}
\end{eqnarray}
These non-perturbative contributions in \(1/\epsilon(q)\) generate the string-induced terms in the real and imaginary parts of the in-medium potential.

\subsection{Real and imaginary parts of the \(Q\bar{Q}\) potential}
\label{complexpotential}

\begin{figure}%
\begin{center}
\includegraphics[width=4.6cm,height=4.8cm]{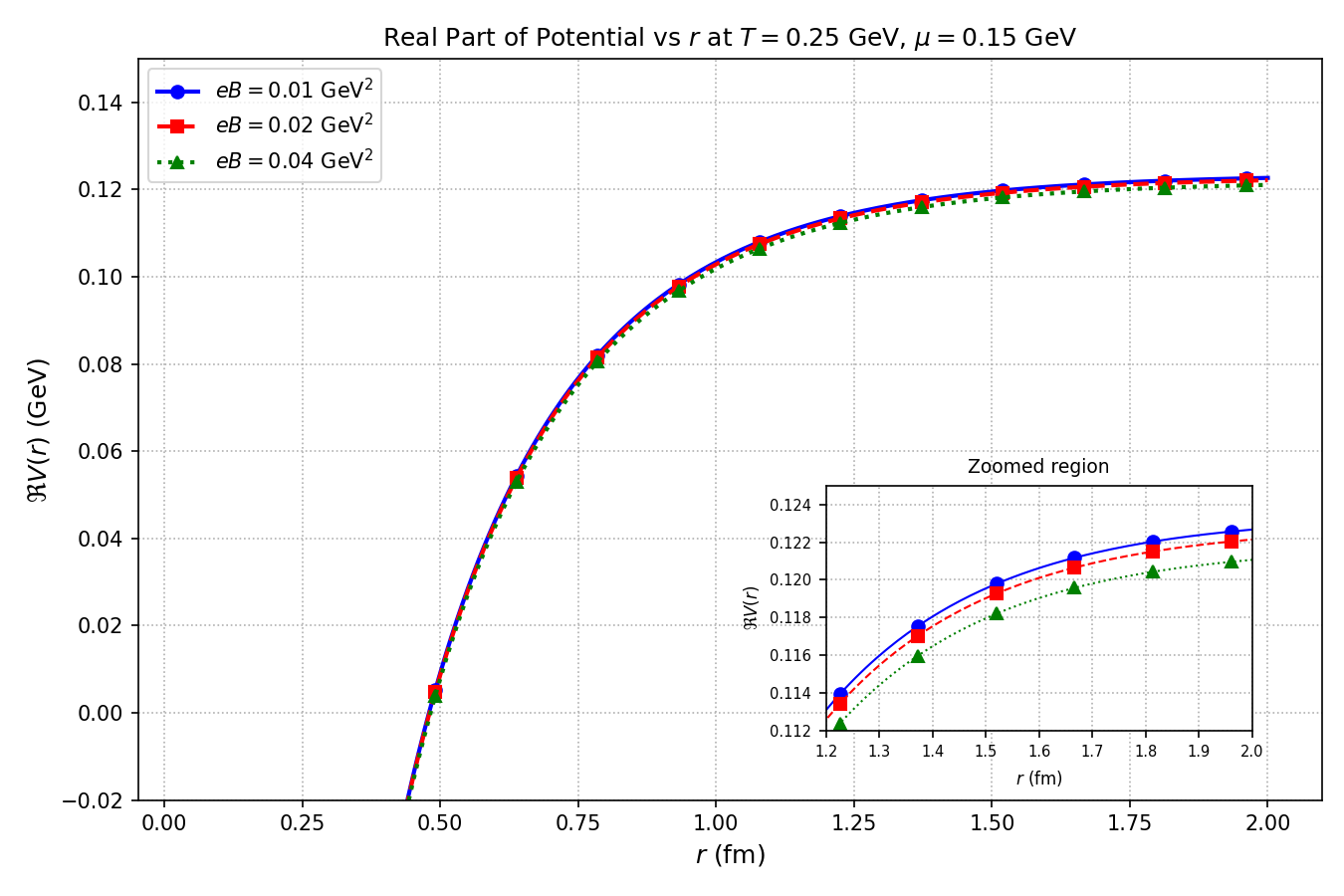}
\includegraphics[width=4.6cm,height=4.8cm]{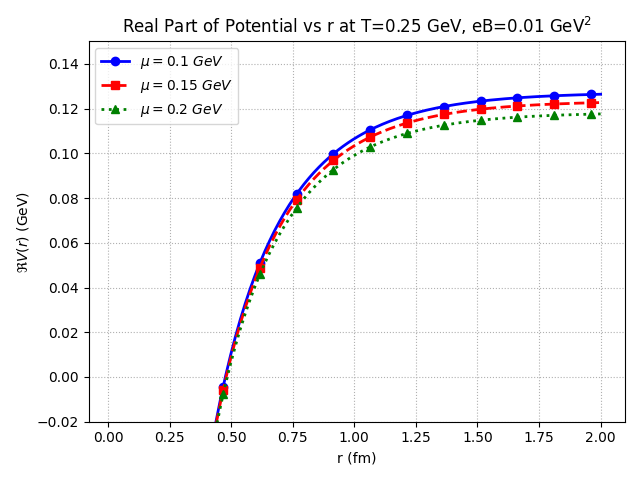}
\includegraphics[width=4.6cm,height=4.8cm]{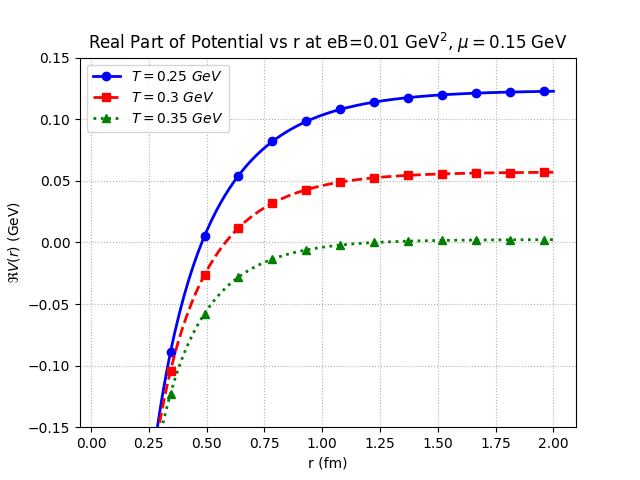}
\caption{Real part of the potential for different strengths of magnetic field (right panel), different strengths of chemical potential (middle panel), and for different strengths of temperature (left panel).}
\label{realb}
\end{center}
\end{figure}

We now compute the real and imaginary parts of the heavy-quark potential in the presence of temperature, chemical potential, and weak magnetic field. Substituting Eq.~\eqref{real_dielectric} into Eq.~\eqref{pot_defn}, one obtains the real part of the in-medium potential (with \(\hat{r}=rm_D\)):
\begin{eqnarray}
{\rm\Re}\,V(r;T,\mu,B)&=&-\frac{4}{3}\alpha_s\left(\frac{e^{-\hat{r}}}{r}+m_D\right)
+\frac{4}{3}\frac{\sigma}{m_D}\left(1-e^{-\hat{r}}\right).
\label{real_potential}
\end{eqnarray}
The dependence on \(T\), \(\mu\), and \(eB\) enters primarily through \(m_D\) (and through \(\alpha_s\) via the chosen running coupling). In Fig.~\ref{realb}, we show \({\rm Re}\,V(r)\) as a function of \(r\).

For the imaginary part, substituting Eq.~\eqref{imaginary_dielectric} into Eq.~\eqref{pot_defn} yields
\begin{eqnarray}
{\rm \Im}\,V_C(r;T,\mu,B)&=&-\frac{4}{3}\frac{\alpha_s T\,M^2_{(T,\mu,B)}}{m_D^2}\,\phi_2(\hat{r}),\\
{\rm \Im}\,V_S(r;T,\mu,B)&=&-\frac{4\sigma T\,M^2_{(T,\mu,B)}}{m_D^4}\,\phi_3(\hat{r}),
\label{imaginary_potential}
\end{eqnarray}
and the total imaginary potential is
\begin{eqnarray}
{\rm \Im}\,V(r;T,\mu,B)={\rm \Im}\,V_C(r;T,\mu,B)+{\rm \Im}\,V_S(r;T,\mu,B).
\end{eqnarray}
The functions \(\phi_2(\hat{r})\) and \(\phi_3(\hat{r})\) are~\cite{Guo:PRD100'2019}
\begin{eqnarray}
\phi_2(\hat{r})&=&2\int_0^{\infty}\frac{z\,dz}{(z^2+1)^2}
\left[1-\frac{\sin(z\hat{r})}{z\hat{r}}\right],\\
\phi_3(\hat{r})&=&2\int_0^{\infty}\frac{z\,dz}{(z^2+1)^3}
\left[1-\frac{\sin(z\hat{r})}{z\hat{r}}\right].
\end{eqnarray}
In the small-\(\hat{r}\) limit (\(\hat{r}\ll 1\)), these functions reduce to
\begin{eqnarray}
\phi_2(\hat{r})&\approx&-\frac{1}{9}\hat{r}^2\left(3\ln \hat{r}-4+3\gamma_E\right),\\
\phi_3(\hat{r})&\approx&-\frac{\hat{r}^2}{12}+\frac{\hat{r}^4}{900}
\left(15\ln\hat{r}-23+15\gamma_E\right).
\end{eqnarray}
Keeping only the leading-log terms in this limit, one finds
\begin{eqnarray}
{\rm \Im}\,V_C(r;T,\mu,B)&=&\frac{4}{9}\frac{\alpha_s T\,M^2_{(T,\mu,B)}}{m_D^2}\,
\hat{r}^2\ln \hat{r},
\label{imaginary_coulomb}\\
{\rm \Im}\,V_S(r;T,\mu,B)&=&-\frac{\sigma T\,M^2_{(T,\mu,B)}}{15m_D^4}\,
\hat{r}^4\ln \hat{r}.
\label{imaginary_string}
\end{eqnarray}

\begin{figure}
\begin{center}
\includegraphics[width=4.6cm,height=4.8cm]{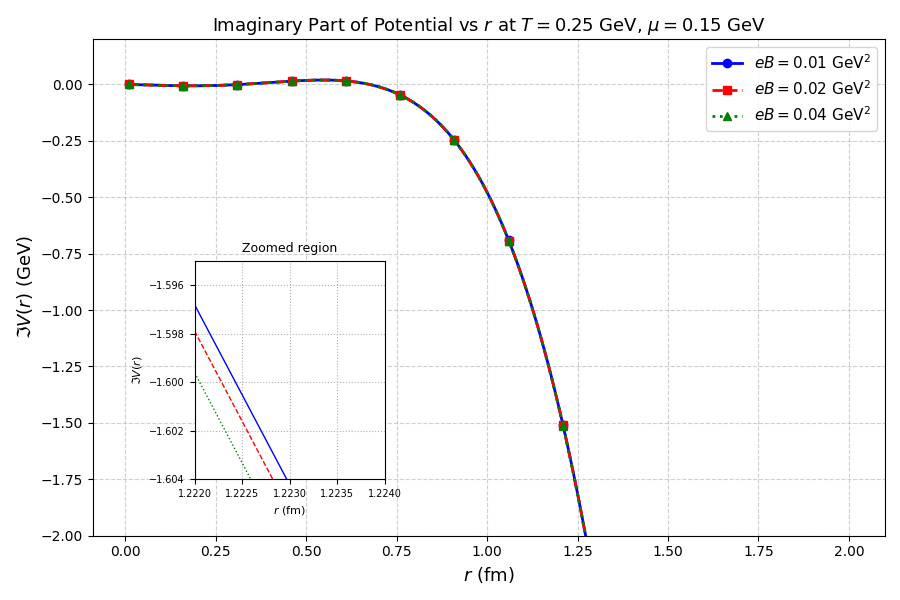}
\includegraphics[width=4.6cm,height=4.8cm]{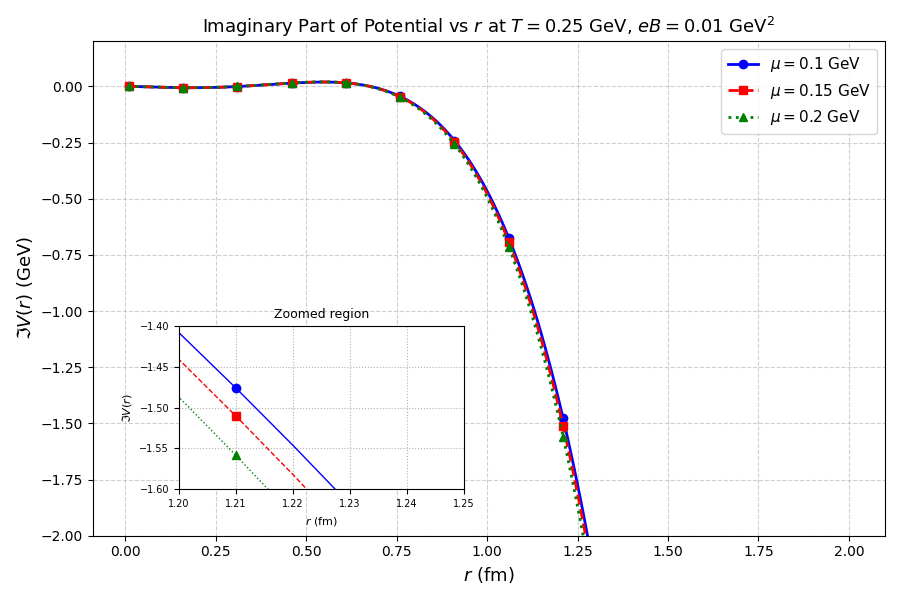}
\includegraphics[width=4.6cm,height=4.8cm]{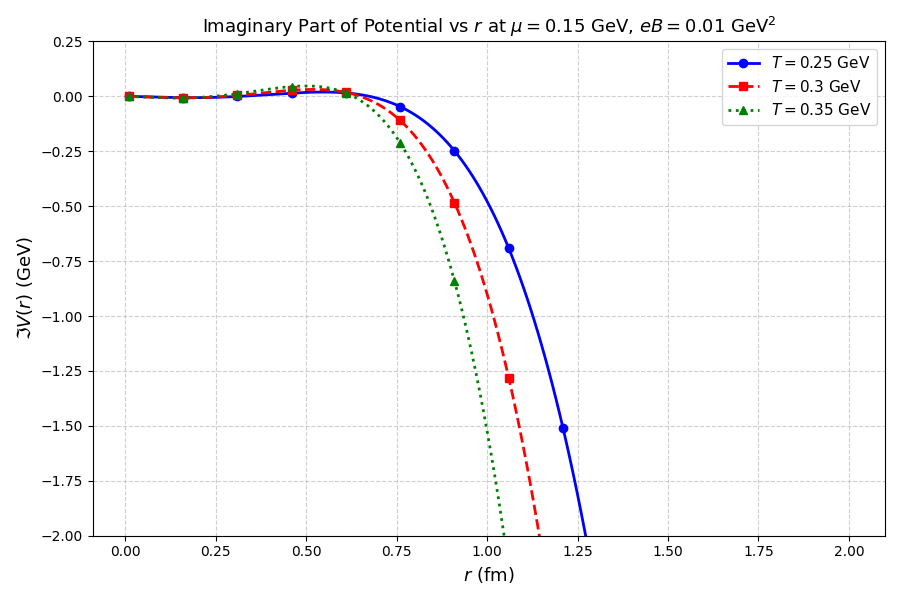}
\caption{Imaginary part of the potential for different strengths of temperature (right panel),
different strengths of chemical potential (middle panel) and for different strengths of magnetic field (left panel).}
\label{imgb}
\end{center}
\end{figure}

In our calculations, \({\rm \Im}\,V\) is treated as a perturbation to the real potential, and the full expressions in terms of \(\phi_2\) and \(\phi_3\) are used when evaluating widths. The small-\(\hat{r}\) limit is useful for understanding parametric behavior and the regime in which the perturbative treatment is justified, i.e., when the typical bound-state size remains smaller than the screening length.
We show \({\rm Im}\,V(r)\) as a function of \(r\) in Fig.~\ref{imgb}.

\section{Properties and Dissociation of Heavy Quarkonia}
\label{properties}

In this section, we use the real and imaginary parts of the in-medium potential to compute binding energies and thermal widths of heavy quarkonia, and then infer their dissociation temperatures in a weakly magnetized thermal medium at finite chemical potential.

\subsection{Binding energy}

\begin{figure}
\begin{center}
\includegraphics[width=8cm,height=8cm]{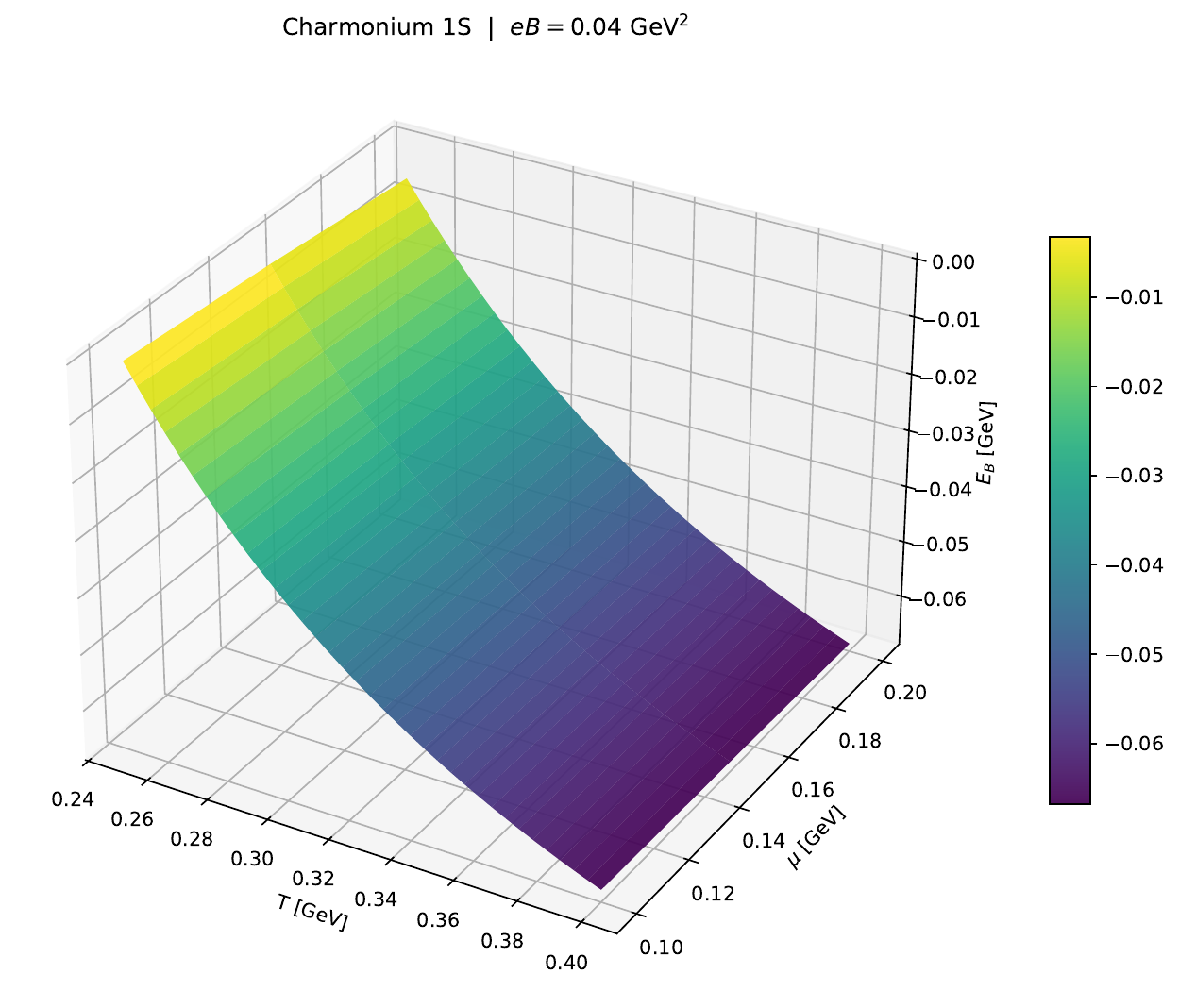}
\includegraphics[width=8cm,height=8cm]{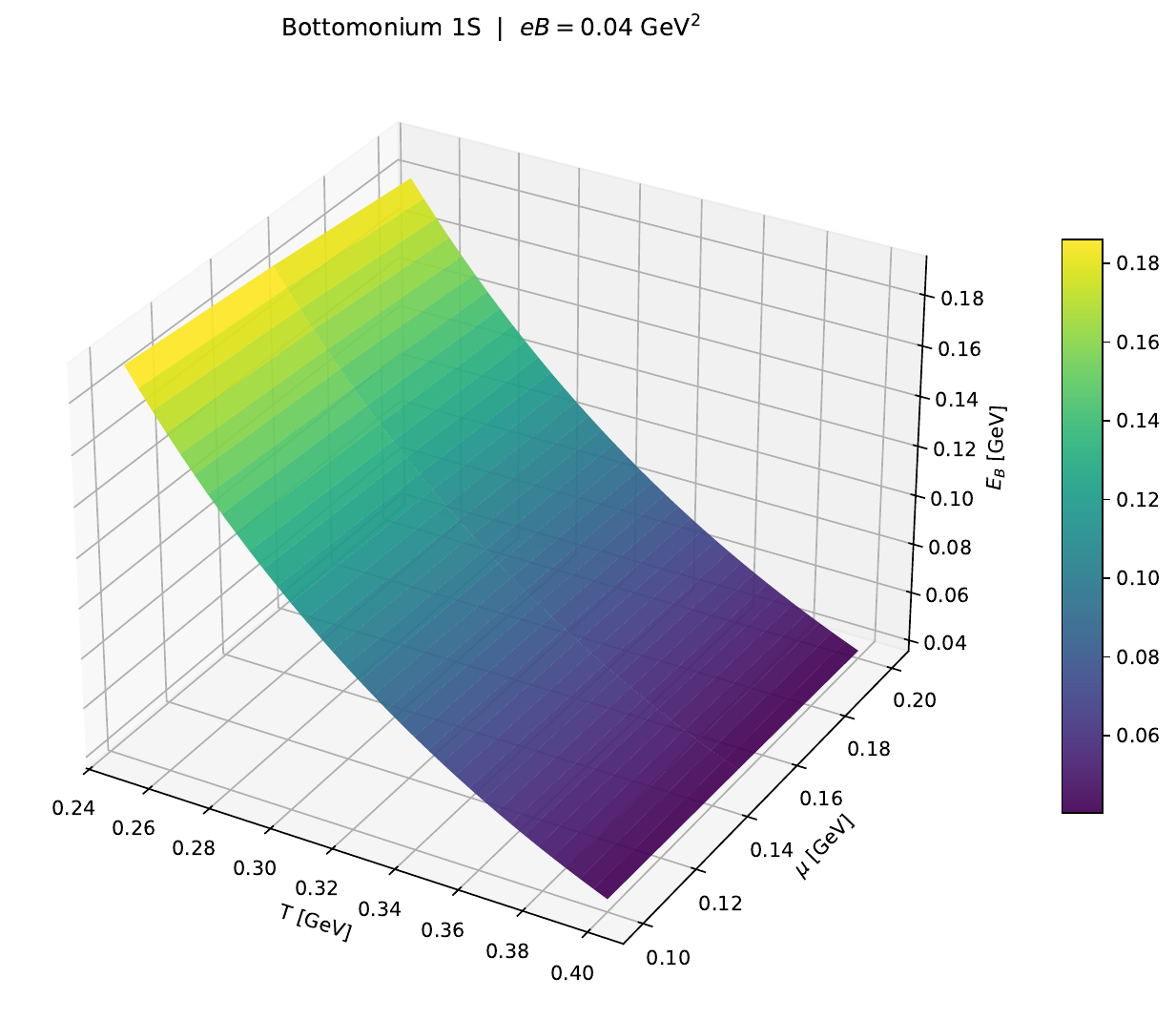}
\caption{Variation of BE for charmonium 1s (left panel) and for bottomium 1s(right panel) at magnetic field $0.04~{\rm GeV}^2$.}
\label{be1}
\end{center}
\end{figure}

\begin{figure}
\begin{center}
\includegraphics[width=8cm,height=8cm]{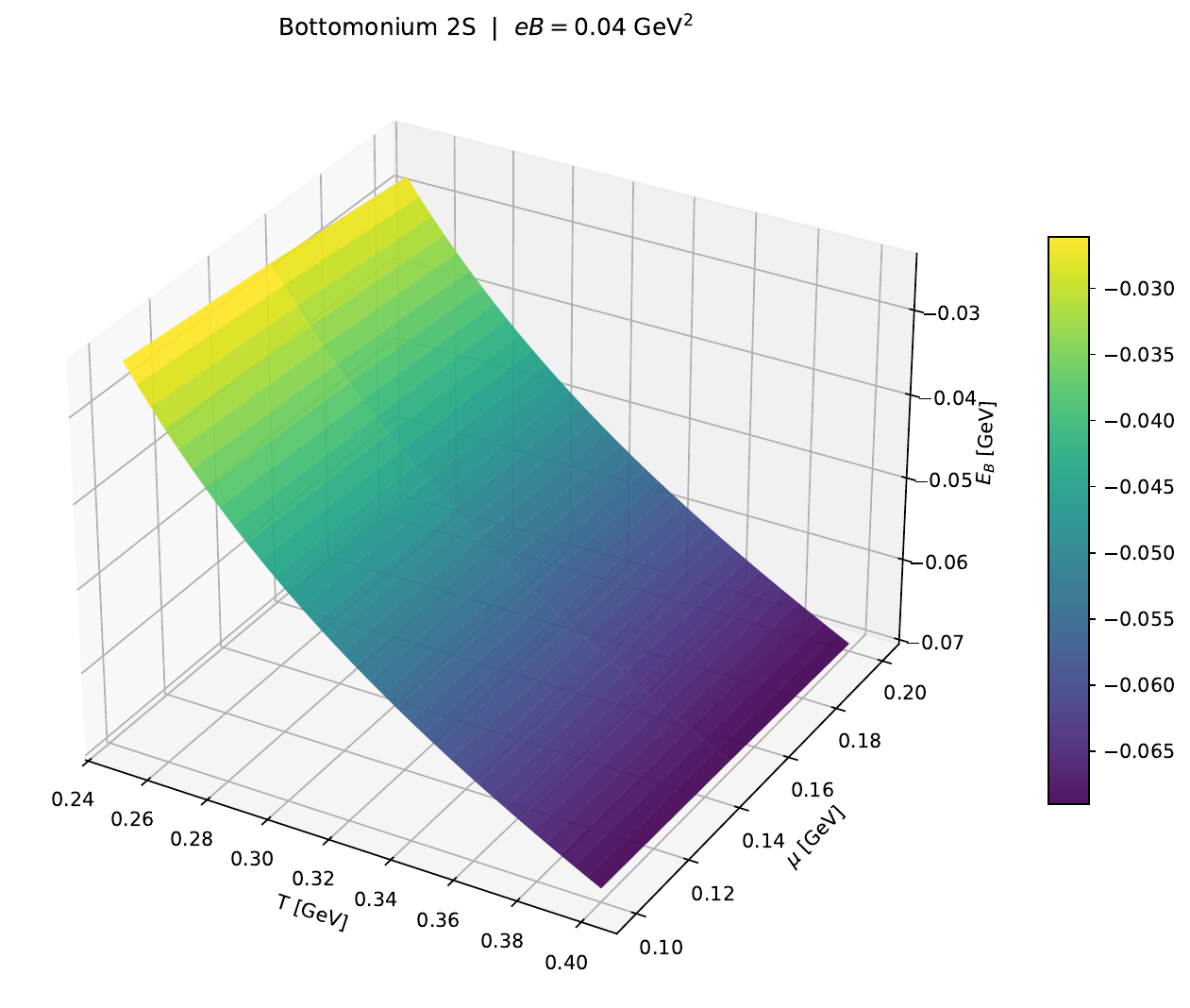}
\caption{Variation of BE for bottomonium 2s at magnetic field $0.04~{\rm GeV}^2$.}
\label{be2}
\end{center}
\end{figure}

We obtain the binding energies and radial wavefunctions by solving the time-independent radial Schr\"odinger equation numerically using the real part of the in-medium potential in Eq.~\eqref{real_potential}. In our numerical results (see the corresponding figures in Sec.~\ref{result}), the binding energies decrease with increasing temperature due to enhanced screening. For representative states, such as \(J/\psi(1S)\), \(\Upsilon(1S)\), and \(\Upsilon(2S)\) at fixed \(eB=0.04~{\rm GeV}^2\), the ground-state bottomonium remains the most strongly bound, while excited states melt earlier. A typical hierarchy is
\[
E_{B}(\Upsilon(1S)) \;>\; E_{B}(\Upsilon(2S)) \;>\; E_{B}(J/\psi(1S)).
\]

\subsection{Thermal width}

\begin{figure}
\begin{center}
\includegraphics[width=8cm,height=8cm]{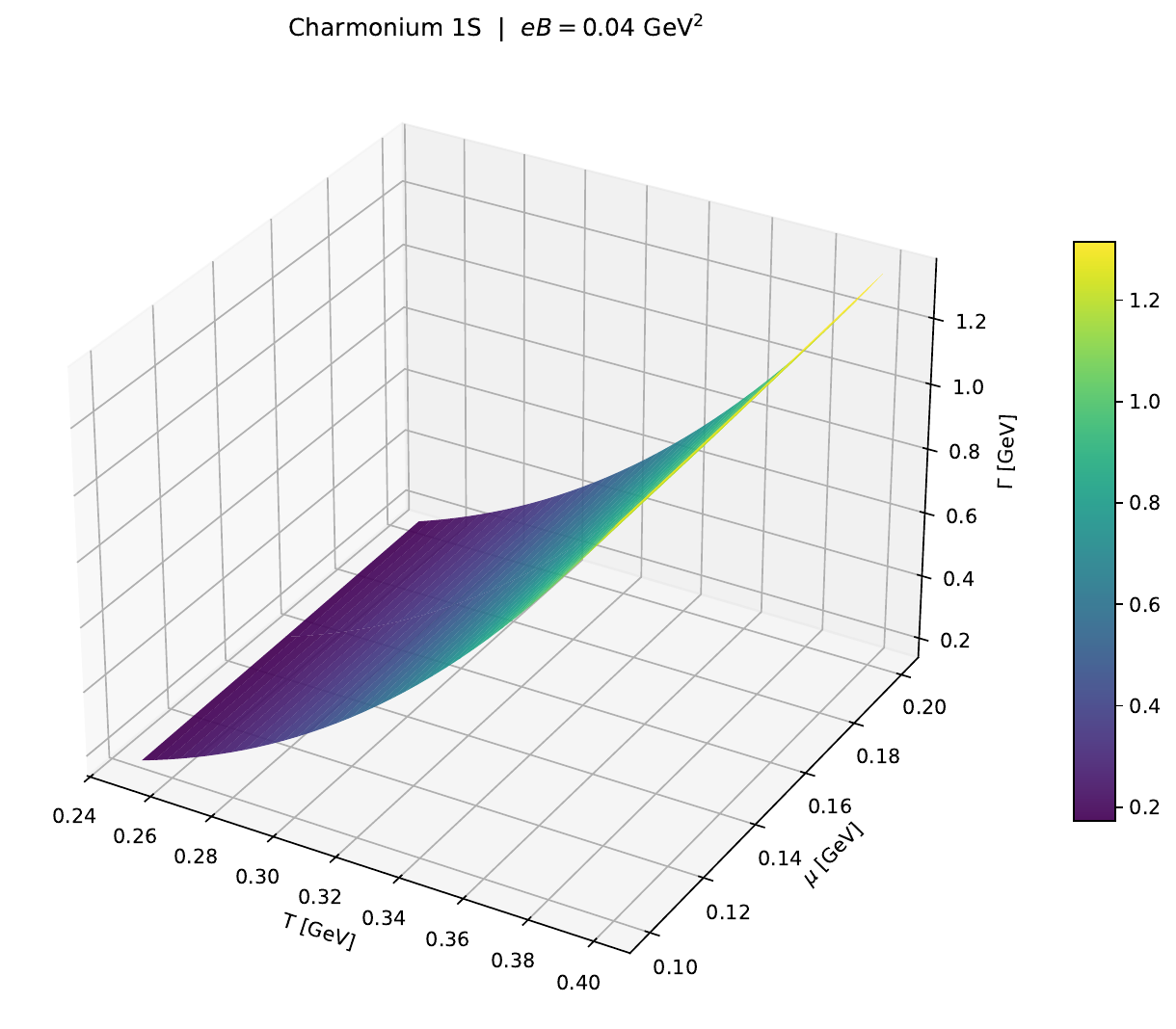}
\includegraphics[width=8cm,height=8cm]{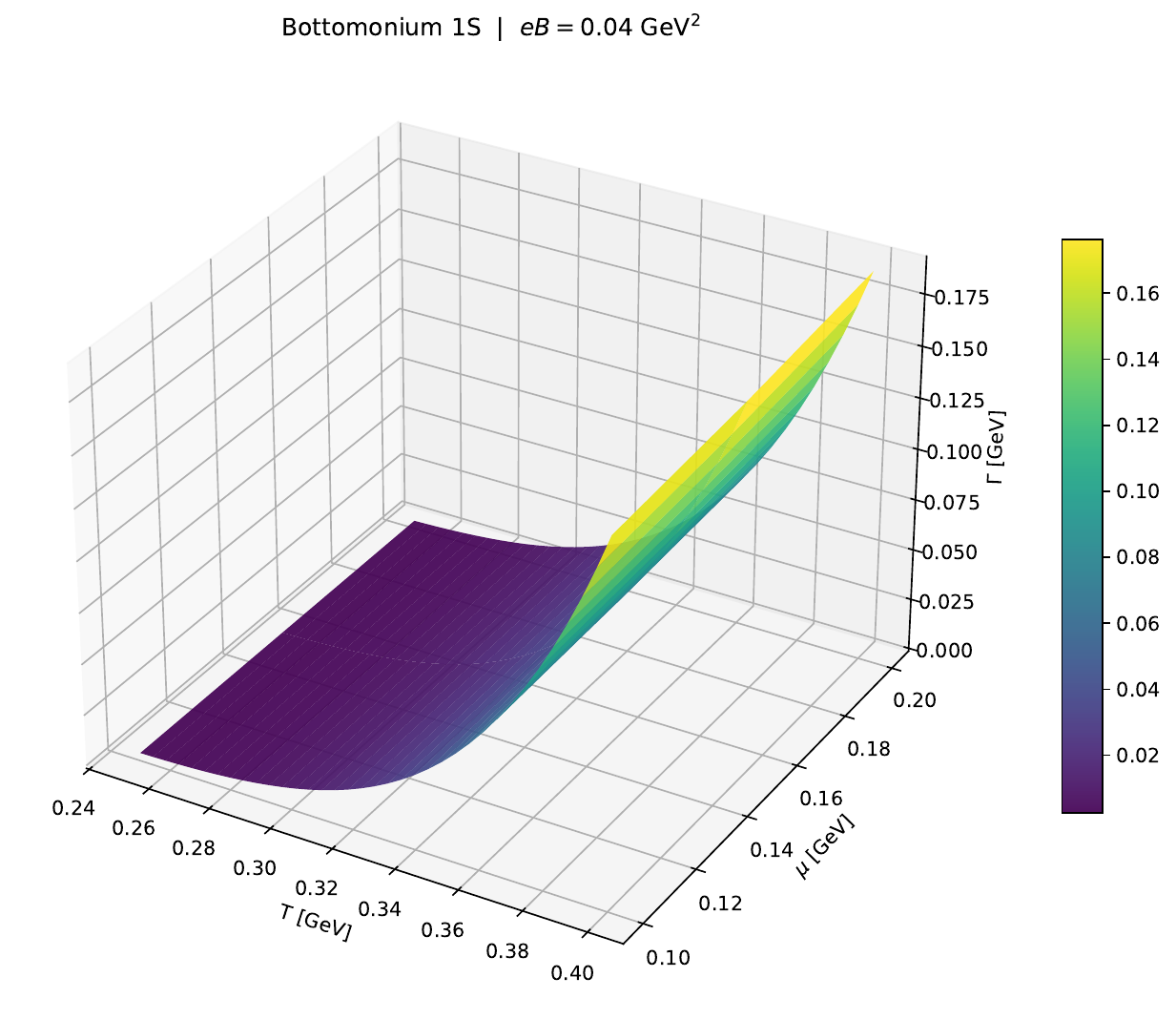}
\caption{Gamma for charmonium 1s (left panel) and for bottomium 1s(right panel) at magnetic field $0.04~\text{GeV}^2$.}
\label{therm1}
\end{center}
\end{figure}

\begin{figure}
\begin{center}
\includegraphics[width=8cm,height=8cm]{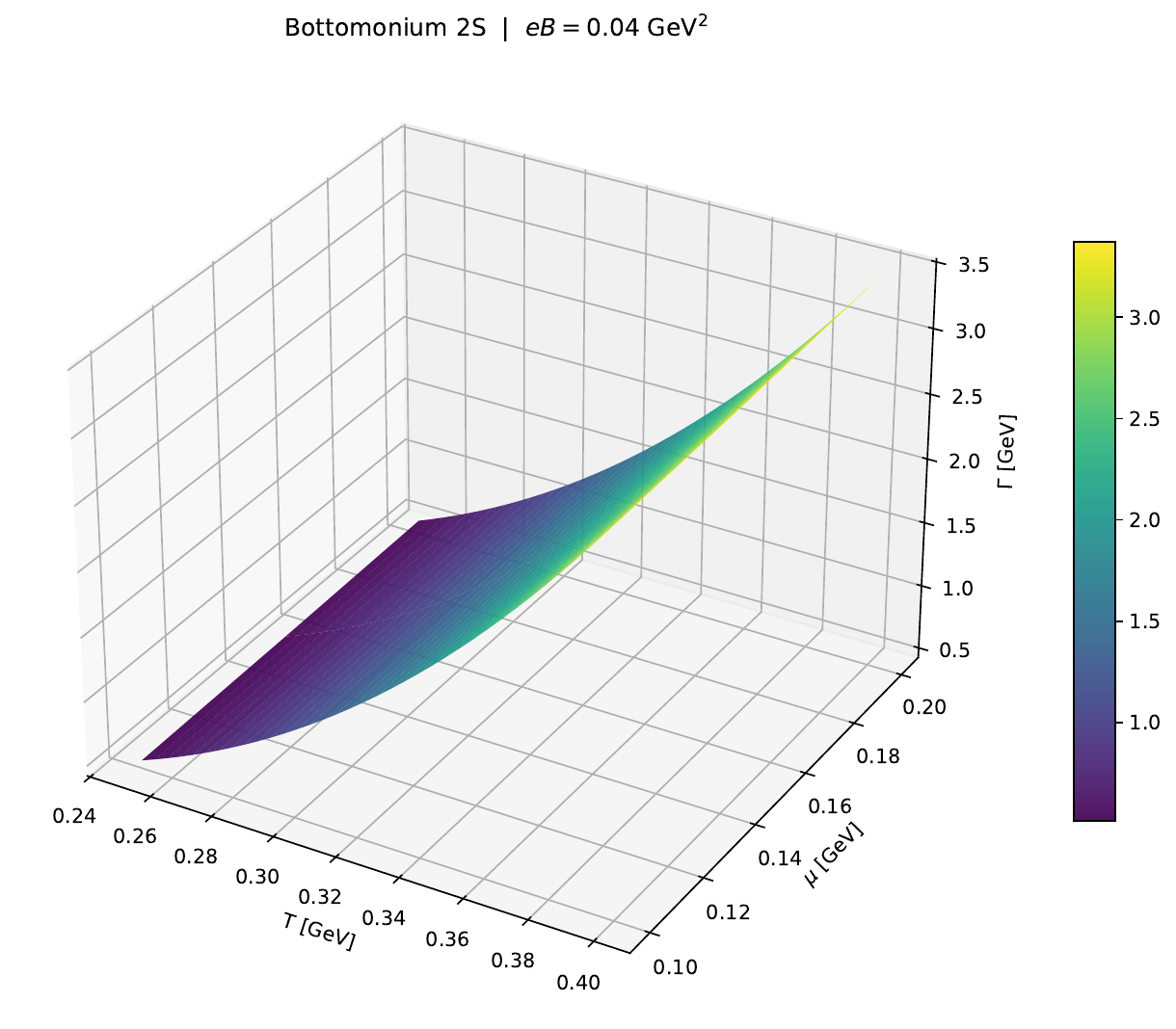}
\caption{Gamma for bottomium 2s at magnetic field $0.04~\text{GeV}^2$.}
\label{therm2}
\end{center}
\end{figure}

The thermal width \(\Gamma\) quantifies the in-medium decay rate (or inverse lifetime) of a quarkonium state due to medium-induced dissociation processes. In a complex-potential approach, the imaginary part acts as a dissipative perturbation and generates a finite width. To leading order in perturbation theory, the width is given by the expectation value of \({\rm Im}\,V\) over the normalized wavefunction:
\begin{eqnarray}
\Gamma(T,\mu,B)=-\int d^3r\;{\rm \Im}\,V(r;T,\mu,B)\,|\Psi(\mathbf{r})|^2.
\label{gammaT}
\end{eqnarray}
For \(S\)-wave states, this can equivalently be written as \(\Gamma=-4\pi\int_0^\infty dr\, r^2\,{\rm \Im}\,V(r)\,|\Psi(r)|^2\), using the corresponding radial normalization.

In the results presented later, \(\Gamma\) increases with temperature, reflecting the strengthening of Landau-damping induced broadening as the medium becomes hotter. States that are more weakly bound develop larger widths and dissociate earlier, while tightly bound states such as \(\Upsilon(1S)\) remain comparatively stable.

\subsection{Medium-induced dissociation criterion}

Having obtained both the binding energy \({\rm BE}\) and the thermal width \(\Gamma\), we estimate dissociation temperatures using the resonance-melting criterion~\cite{Mocsy:PRL99'2007}
\[
\Gamma \ge 2\,{\rm BE}.
\]
This criterion encodes the condition that the in-medium width becomes comparable to (or larger than) the energy required to keep the state bound, signalling the disappearance of a well-defined quarkonium resonance.

\begin{table}[h!]
\centering
\begin{tabular}{|c|c|c|c|c|}
\hline
\multicolumn{5}{|c|}{$T_d$~(GeV)} \\ \hline
$eB$~($GeV^2$) & $\mu$~(GeV) & $J/\psi$ & $\Upsilon$ & $\Upsilon^{\prime}$ \\ \hline
\multirow{3}{*}{0.01} 
& 0.1 &  0.253 & 0.379& 0.254 \\ \cline{2-5}
& 0.15 & 0.250 & 0.377 & 0.250 \\ \cline{2-5}
& 0.2 &  0.248&  0.375&  0.247\\ \hline
\multirow{3}{*}{0.04} 
& 0.1 & 0.252 & 0.378 & 0.252 \\ \cline{2-5}
& 0.15 & 0.249 & 0.376 & 0.250 \\ \cline{2-5}
& 0.2 & 0.247 & 0.373& 0.246  \\ \hline
\end{tabular}
\caption{Dissociation temperatures $T_d$ of quarkonium states as a function of chemical potential $\mu$ and magnetic field $eB$.}
\label{tab}
\end{table}

\section{Results}
\label{result}

In this section, we present and discuss our numerical results for the heavy-quark potential and quarkonium properties in the presence of a weak magnetic field \(eB\) and finite chemical potential \(\mu\).

\subsection{Running coupling}

Figure~\ref{alphat} shows the dependence of the strong coupling \(\alpha_s\) on temperature and magnetic field, using the running prescription introduced earlier. The left panel displays \(\alpha_s\) as a function of \(eB\) at fixed temperature \(T = 0.25~\mathrm{GeV}\) and chemical potential \(\mu = 0.200~\mathrm{GeV}\), while the right panel shows the temperature dependence at fixed \(eB = 0.02~\mathrm{GeV}^2\).

The coupling decreases monotonically with increasing temperature, dropping from \(\alpha_s\simeq 0.315\) at \(T=0.25~\mathrm{GeV}\) to \(\alpha_s\simeq 0.262\) at \(T=0.40~\mathrm{GeV}\). This trend follows directly from the one-loop logarithmic running, since the scale \(\Lambda\) increases with \(T\), leading to a suppression of \(\alpha_s\) consistent with asymptotic freedom. In contrast, \(\alpha_s\) shows a slight increasing trend with magnetic field: it increases marginally from \(\alpha_s=0.31470\) at \(eB=0.01~\mathrm{GeV}^2\) to \(\alpha_s=0.31496\) at \(eB=0.02~\mathrm{GeV}^2\). Thus, temperature and magnetic field act in opposite directions: increasing \(T\) reduces \(\alpha_s\) significantly, whereas \(eB\) produces only a mild enhancement. This interplay influences screening through quantities such as the Debye mass and ultimately affects quarkonium binding and thermal widths, as discussed below.

\subsection{Debye screening mass}

Figure~\ref{debye} summarizes the behavior of the Debye mass \(m_D\) in the parameter range explored. The left panel shows \(m_D\) as a function of magnetic field \(eB\) at fixed \(T = 0.25~\mathrm{GeV}\) for three values of chemical potential \(\mu = 0.100,\,0.150,\) and \(0.200~\mathrm{GeV}\). A slight increase in \(m_D\) is observed with increasing \(eB\), indicating that a magnetic background marginally enhances screening. For each \(eB\), \(m_D\) also increases with \(\mu\), such that
\[
m_D(\mu = 0.200~\mathrm{GeV}) > m_D(\mu = 0.150~\mathrm{GeV}) > m_D(\mu = 0.100~\mathrm{GeV}).
\]
The middle panel shows \(m_D\) as a function of \(\mu\) at fixed \(eB = 0.02~\mathrm{GeV}^2\) for \(T = 0.25,\,0.30,\) and \(0.40~\mathrm{GeV}\). The increase with \(\mu\) is mild, whereas the temperature dependence is much stronger: \(m_D\) grows substantially as \(T\) rises, which is evident from the well-separated curves. The right panel displays \(m_D\) versus \(T\) at fixed \(\mu = 0.100~\mathrm{GeV}\) for \(eB = 0.01,\,0.02,\) and \(0.04~\mathrm{GeV}^2\). The Debye mass increases almost linearly with temperature, implying stronger color screening at higher \(T\). The magnetic-field effect remains comparatively weak, although a small enhancement of \(m_D\) with increasing \(eB\) is visible.

This pattern is consistent with the analytic structure of \(m_D^2(T,\mu,B)\) in Eq.~\ref{debyemass}. The leading thermal contribution,
\[
\frac{g^2 T^2}{3}\Big(N_c+\tfrac{N_f}{2}\Big),
\]
scales as \(T^2\) and therefore dominates the temperature dependence, accounting for the strong rise of \(m_D\) with \(T\). The chemical-potential dependence enters through the \(6N_f g^2\hat{\mu}^2\) term; since \(\hat{\mu}=\mu/(2\pi T)\) is small in the considered range, its net effect is a modest increase in \(m_D\). The magnetic-field corrections enter at order \((q_f B)^2\) and are suppressed by explicit factors of \(1/T^2\) and by the Bessel-function sums, which decrease rapidly when \(m_f l/T \gtrsim 1\). Consequently, the net \(eB\) dependence remains weak for the moderate temperatures explored. Altogether, the sensitivity hierarchy observed numerically can be summarized as
\[
T \;>\; \mu \;>\; eB,
\]
i.e., temperature is the principal driver of Debye screening, while finite chemical potential and magnetic field provide smaller corrections in this parameter window.

\subsection{Real part of the potential}

Figure~\ref{realb} shows the variation of the real part of the heavy-quark potential, \(\Re[V(r)]\), with inter-quark separation \(r\) for different values of \(eB\), \(\mu\), and \(T\). The left panel presents \(\Re[V(r)]\) for three magnetic fields \(eB = 0.01,\,0.02,\) and \(0.04~\mathrm{GeV}^2\) at fixed \(T = 0.25~\mathrm{GeV}\) and \(\mu = 0.15~\mathrm{GeV}\). The potential decreases mildly with increasing magnetic field, which is also highlighted in the zoomed region, indicating a small enhancement of effective screening in a magnetic background. The middle panel shows \(\Re[V(r)]\) for \(\mu = 0.10,\,0.15,\) and \(0.20~\mathrm{GeV}\) at fixed \(T = 0.25~\mathrm{GeV}\) and \(eB = 0.01~\mathrm{GeV}^2\). A slight suppression at larger separations is observed as \(\mu\) increases, again consistent with the mild growth of \(m_D\) at finite density. The right panel shows \(\Re[V(r)]\) for \(T = 0.25,\,0.30,\) and \(0.35~\mathrm{GeV}\) at fixed \(eB = 0.01~\mathrm{GeV}^2\) and \(\mu = 0.15~\mathrm{GeV}\). The temperature dependence is the strongest: as \(T\) increases, the magnitude of the potential decreases across all \(r\), reflecting the rapid increase in the Debye screening mass and the corresponding flattening of the potential at shorter distances. Overall, the relative impact follows the hierarchy
\[
T \;>\; \mu \;>\; eB,
\]
so that thermal screening dominates over density and magnetic-field effects in the present setup.

\subsection{Imaginary part of the potential}

The behavior of the imaginary part, \(\Im V(r)\), is shown in Fig.~\ref{imgb}. The left panel shows \(\Im V(r)\) for different magnetic fields \(eB = 0.01,\,0.02,\) and \(0.04~\text{GeV}^2\) at fixed \(T = 0.25~\text{GeV}\) and \(\mu = 0.15~\text{GeV}\). The curves nearly overlap over most of the \(r\) range; the zoomed region indicates that \(|\Im V(r)|\) decreases slightly with increasing \(eB\), implying a very mild suppression of the imaginary part in a stronger magnetic background within the explored regime.

The middle panel shows \(\Im V(r)\) for \(\mu = 0.10,\,0.15,\) and \(0.20~\text{GeV}\) at fixed \(T = 0.25~\text{GeV}\) and \(eB = 0.01~\text{GeV}^2\). The dependence on \(\mu\) is again weak: the inset highlights that the imaginary part becomes slightly less negative as \(\mu\) increases, corresponding to a marginal reduction in damping.

The right panel shows \(\Im V(r)\) for \(T = 0.25,\,0.30,\) and \(0.35~\text{GeV}\) at fixed \(\mu = 0.15~\text{GeV}\) and \(eB = 0.01~\text{GeV}^2\). For all temperatures, \(\Im V(r)\) is close to zero at small separations (\(r < 0.5~\text{fm}\)), while it becomes more negative at larger \(r\), reflecting enhanced Landau damping and thermal broadening. The magnitude grows strongly with temperature; for example, around \(r \approx 1.2~\text{fm}\), \(\Im V(r)\) reaches approximately \(-1.5~\text{GeV}\), \(-1.75~\text{GeV}\), and below for \(T = 0.25,\,0.30,\) and \(0.35~\text{GeV}\), respectively. Hence, the dominant control of \(\Im V(r)\) is again provided by temperature, with subleading corrections from \(\mu\) and \(eB\), consistent with the screening hierarchy discussed above.

\subsection{Binding energies}

Figures~\ref{be1} and \ref{be2} show the variation of the binding energy (BE) with temperature for the charmonium \(1S\), bottomonium \(1S\), and bottomonium \(2S\) states at fixed \(eB = 0.04~\text{GeV}^2\). The left panel of Fig.~\ref{be1} shows that the binding energy of the charmonium \(1S\) state decreases with temperature, starting near \(0.06~\text{GeV}\) at \(T \approx 0.24~\text{GeV}\) and dropping to around \(0.01~\text{GeV}\) at \(T \approx 0.40~\text{GeV}\). The right panel of Fig.~\ref{be1} shows that the bottomonium \(1S\) state remains more strongly bound, with BE decreasing from about \(0.18~\text{GeV}\) at \(T = 0.24~\text{GeV}\) to \(0.06~\text{GeV}\) at \(T = 0.40~\text{GeV}\). Figure~\ref{be2} shows that the bottomonium \(2S\) state is more weakly bound: its BE decreases from about \(0.065~\text{GeV}\) at \(T = 0.24~\text{GeV}\) to around \(0.03~\text{GeV}\) at \(T = 0.40~\text{GeV}\), indicating earlier melting of the excited state.

A clear hierarchy of binding strengths emerges:
\[
E_B(\Upsilon(1S)) > E_B(\Upsilon(2S)) > E_B(J/\psi(1S)).
\]
The stronger binding of bottomonium relative to charmonium reflects the heavier quark mass and the reduced sensitivity to screening. In the present parameter set, magnetic-field effects at \(eB = 0.04~\text{GeV}^2\) provide only modest modifications compared to the dominant thermal suppression.

\subsection{Thermal widths}

The thermal decay widths \(\Gamma\) as functions of temperature for different quarkonium states at fixed \(eB = 0.04~\text{GeV}^2\) are shown in Figs.~\ref{therm1} and \ref{therm2}. The left panel of Fig.~\ref{therm1} shows that the width of the charmonium \(1S\) state increases monotonically with temperature, rising from approximately \(0.2~\text{GeV}\) at \(T = 0.24~\text{GeV}\) to about \(1.2~\text{GeV}\) at \(T = 0.40~\text{GeV}\). This strong growth signals substantial in-medium broadening and rapid destabilization of the \(J/\psi\) state as temperature increases. The right panel of Fig.~\ref{therm1} shows that the bottomonium \(1S\) state has a much smaller width, increasing from about \(0.02~\text{GeV}\) at \(T = 0.24~\text{GeV}\) to \(0.16~\text{GeV}\) at \(T = 0.40~\text{GeV}\), consistent with its stronger binding and greater thermal stability.

Figure~\ref{therm2} shows the bottomonium \(2S\) state, whose width increases from roughly \(1.0~\text{GeV}\) at \(T = 0.24~\text{GeV}\) to nearly \(3.0~\text{GeV}\) at \(T = 0.40~\text{GeV}\). The rapid broadening indicates that the excited \(2S\) state is considerably more vulnerable to thermal fluctuations and dissociates more readily than the ground-state bottomonium.

The ordering of widths is
\[
\Gamma(\Upsilon(2S)) > \Gamma(J/\psi(1S)) > \Gamma(\Upsilon(1S)),
\]
and it is inversely correlated with the binding-energy hierarchy. States with weaker binding develop larger widths and melt earlier, while tightly bound states such as \(\Upsilon(1S)\) remain more resistant to medium-induced suppression.

\subsection{Dissociation temperatures}

Table~\ref{tab} lists the dissociation temperatures \(T_d\) of \(J/\psi\), \(\Upsilon(1S)\), and \(\Upsilon^{\prime}(2S)\) as functions of \(\mu\) and \(eB\). The results exhibit a clear sequential pattern: \(\Upsilon(1S)\) has the highest \(T_d\), followed by \(\Upsilon^{\prime}(2S)\), while \(J/\psi\) dissociates at the lowest temperature. At fixed magnetic field, \(T_d\) decreases mildly with increasing \(\mu\). For example, at \(eB = 0.01~\text{GeV}^2\), the dissociation temperature of \(J/\psi\) decreases from \(0.253~\text{GeV}\) at \(\mu = 0.10~\text{GeV}\) to \(0.248~\text{GeV}\) at \(\mu = 0.20~\text{GeV}\), with similar trends for bottomonium states. This behavior reflects the modest enhancement of screening at higher density.

The magnetic-field effect on \(T_d\) is also weak but systematic: comparing \(eB = 0.01~\text{GeV}^2\) and \(eB = 0.04~\text{GeV}^2\), the dissociation temperatures decrease slightly for all states. Overall, the ordering remains
\[
T_d(\Upsilon(1S)) > T_d(\Upsilon^{\prime}(2S)) > T_d(J/\psi),
\]
supporting the picture of hierarchical quarkonium melting in which more strongly bound states survive to higher temperatures and stronger medium conditions.

\section{Conclusions}
\label{conclusion}

In this work, we have examined how finite quark chemical potential and weak magnetic fields modify the in-medium properties and dissociation of heavy quarkonium states in a hot QCD medium. Starting from the one-loop resummed gluon propagator in the imaginary-time formalism, we incorporated non-perturbative effects through a phenomenological correction to the HTL propagator. This construction yields a complex dielectric permittivity and, in turn, a complex heavy-quark potential. The binding energies of charmonium and bottomonium states were obtained by solving the Schr\"odinger equation with the real part of the potential, while the imaginary part provided the corresponding thermal decay widths. Combining these two ingredients, we extracted dissociation temperatures for different quarkonium states in a magnetized and baryon-rich environment.

A consistent hierarchy emerges in the influence of the medium parameters. Temperature is the dominant driver of screening and quarkonium stability: the Debye mass increases strongly with \(T\), whereas its dependence on chemical potential is mild and its dependence on magnetic field is minimal within the weak-field regime considered. This follows from the analytic structure of the screening mass, in which the leading thermal term is parametrically larger than the finite-density and \((q_fB)^2\) corrections. As a result, the real part of the inter-quark potential is most strongly suppressed by increasing temperature, leading to a marked reduction in binding. Finite chemical potential introduces a smaller but visible additional suppression, while the magnetic field produces only marginal changes to the potential profile in the explored parameter window.

The imaginary part of the potential, which encodes Landau-damping--induced thermal broadening, exhibits the same ordering but with a more pronounced sensitivity to temperature. Its magnitude grows with \(T\), reflecting stronger thermal fluctuations and faster in-medium dissociation at higher temperatures. In comparison, the effects of \(\mu\) and \(eB\) remain subleading and lead only to small quantitative changes in the width. Taken together, these trends suggest that thermal effects primarily govern both the binding and the lifetime of quarkonium states in the plasma.

The computed binding energies for \(J/\psi(1S)\), \(\Upsilon(1S)\), and \(\Upsilon(2S)\) decrease monotonically with temperature, with the bottomonium ground state remaining the most strongly bound. The thermal widths show the opposite trend: more weakly bound states acquire larger widths. This establishes the hierarchy
\begin{equation}
E_{B}(\Upsilon(1S)) > E_{B}(\Upsilon(2S)) > E_{B}(J/\psi),
\end{equation}
\begin{equation}
\Gamma(\Upsilon(2S)) > \Gamma(J/\psi) > \Gamma(\Upsilon(1S)),
\end{equation}
which is consistent with the relative robustness of these states against medium effects. The dissociation temperatures inferred from the competition between binding energy and thermal width follow the expected sequential suppression pattern,
\begin{equation}
T_{d}(\Upsilon(1S)) > T_{d}(\Upsilon(2S)) > T_{d}(J/\psi).
\end{equation}
We find that \(T_d\) decreases mildly with increasing chemical potential and magnetic field strength, indicating enhanced screening in a denser and weakly magnetized medium. Nevertheless, these shifts remain modest compared to the dominant thermal effect.

Overall, our results show that finite density and weak magnetic fields introduce quantitative corrections to the in-medium quarkonium potential and related observables, while temperature remains the primary factor governing quarkonium dissociation in a hot QCD medium. These findings are relevant for interpreting quarkonium suppression patterns in heavy-ion collisions, where temperature, baryon density, and magnetic fields can coexist. A natural extension of this work would involve stronger magnetic fields, more general anisotropic screening, and real-time evolution—key steps toward a complete dynamical description of quarkonium in extreme QCD matter. Furthermore, applying machine learning techniques \cite{Jamal:2025gjy} could significantly refine these results.

\section*{Acknowledgements}
IN acknowledges the Women Scientist Scheme A (WoS A) of the Department of Science and Technology (DST) for funding with grant no. DST/WoS-A/PM-79/2021. M. Y. Jamal would like to acknowledge the Key Laboratory of Quark and Lepton Physics (MOE) \& Institute of Particle Physics, Central China Normal University, Wuhan, China, for providing the postdoctoral fellow position during the tenure of this work.
\appendix
\appendixpage
\addappheadtotoc
In the following appendices, we have shown the explicit 
calculations of form factors $b_0(Q)$ and $b_2(Q)$ from the gluon self-energy in the presence of temperature, weak magnetic field, and chemical potential.
\begin{appendices}
\section{Gluon self energy in a weak magnetic field in the presence of temperature and chemical potential}
For the evaluation of the gluon self-energy, one must consider contributions arising from both the quark loop and the gluon loop in a thermal medium characterized by a weak magnetic field and finite chemical potential. In the presence of a weak magnetic field and finite chemical potential, medium effects enter exclusively through the quark loop contribution, leaving the gluonic contribution unchanged. In an earlier work, one of us computed the gluon self-energy in a weak magnetic field at finite temperature~\cite{Mujeeb:PRD102'2020}. Here, we extend that framework to include the effects of a finite chemical potential. Accordingly, we first focus on the quark-loop contribution to the gluon self-energy. 
\begin{eqnarray}
i\Pi^{\mu\nu}_{ab}(Q)&=&\sum_f\frac{g^2\delta_{ab}}{2}\int\frac{d^4K}{(2\pi)^4}Tr
\left[ \gamma^\nu i S(K) \gamma^\mu i S(P)\right].
\label{self_energy}
\end{eqnarray}
The $S(k)$ is the quark propagator in a weak magnetic field
which can be written up to order of $O(q_fB)^2$ as
\cite{Ayala:REVI66'2020, Chyi:PRD62'2000, Ayala:PRD71'2005} 
\begin{eqnarray}
iS(K)=i\frac{(\slashed{K}+m_f)}{K^2-m^2_f}-q_fB
\frac{\gamma_1\gamma_2(\slashed{K}+m_f)}{(K^2-m_f^2)^2}
-2i(q_f B)^2\frac{[K_\perp^2(\slashed{K}_\parallel+m_f)
+\slashed{K}_\perp(m_f^2-K_\parallel^2)]}
{(K^2-m_f^2)^4},
\label{weak_propagator}
\end{eqnarray}
where $m_f$ and $q_f$ are the mass and charge of the 
$f^{th}$ flavor quark.
After simplifying, the above gluon self-energy given by 
Eq.\eqref{self_energy} can be expressed as follows
\begin{eqnarray}
\Pi^{\mu\nu}(Q)=\Pi^{\mu\nu}_{(0,0)}(Q)+\Pi^{\mu\nu}_{(1,1)}(Q)
+2\Pi^{\mu\nu}_{(2,0)}(Q)+O[(q_fB)^3],
\label{self_energy2}
\end{eqnarray}
\begin{eqnarray}
\Pi^{\mu\nu}_{(0,0)}(Q)
&=&\sum_f i2g^2\int\frac{d^4K}{(2\pi)^4}\frac{\left[P^\mu K^\nu+
K^\mu P^\nu-g^{\mu\nu}(K.P-m_f^2)\right]}
{(K^2-m^2_f)(P^2-m_f^2)},\\
\Pi^{\mu\nu}_{(1,1)}(Q)
&=&\sum_f 2ig^2(q_fB)^2\int\frac{d^4K}{(2\pi)^4}
\frac{\left[P_\parallel^\mu K_\parallel^\nu +K_\parallel^\mu 
P_\parallel^\nu +(g_\parallel^{\mu\nu}-g_\perp^{\mu\nu})
(m_f^2-K_\parallel .P_\parallel)\right]}
{(K^2-m^2_f)^2(P^2-m_f^2)^2},\\
\Pi^{\mu\nu}_{(2,0)}(Q)
&=&-\sum_f 4ig^2(q_fB)^2\int\frac{d^4K}{(2\pi)^4}
\left[\frac{M^{\mu\nu}}{(K^2-m_f^2)^4(P^2-m_f^2)}\right],
\end{eqnarray}
where
\begin{eqnarray}
M^{\mu\nu}&=&K_\perp^2\left[P^\mu K_\parallel^\nu+
K_\parallel^\mu P^\nu-g^{\mu\nu}(K_\parallel .P-m_f^2)\right]
+(m_f^2-K_\parallel^2)\left[P^\mu K_\perp^\nu +K_\perp^\mu 
P^\nu-g^{\mu\nu}(K_\perp .P)\right].\nonumber\\
\end{eqnarray}

\noindent Before proceeding further, we first discuss the general structure of the gluon self-energy in a thermal medium in the presence of a weak magnetic field along $z$-direction, represented by the vector $n_\mu=(0,0,0,-1)$.  In a thermal medium, the heat bath defines a preferred rest frame $u^\mu=(1,0,0,0)$, thus, the gluon self-energy can be expressed in terms of the available four-vectors and tensors consistent with the symmetries of the system~\cite{karmakar:EPJC79'2019} 
\begin{eqnarray}
\Pi^{\mu\nu}(Q)=b(Q)B^{\mu\nu}(Q)+c(Q)R^{\mu\nu}(Q)+d(Q)M^{\mu\nu}(Q)
+a(Q)N^{\mu\nu}(Q),
\label{self_decomposition}
\end{eqnarray}
where 
\begin{eqnarray}
B^{\mu\nu}(Q)&=&\frac{{\bar{u}}^\mu{\bar{u}}^\nu}{{\bar{u}}^2},\\
R^{\mu\nu}(Q)&=&g_{\perp}^{\mu\nu}-\frac{Q_{\perp}^{\mu}Q_{\perp}^{\nu}}
{Q_{\perp}^2},\\
M^{\mu\nu}(Q)&=&\frac{{\bar{n}}^\mu{\bar{n}}^\nu}{{\bar{n}}^2},\\
N^{\mu\nu}(Q)&=&\frac{{\bar{u}}^\mu{\bar{n}}^\nu+{\bar{u}}^\nu{\bar{n}}^\mu}
{\sqrt{{\bar{u}}^2}\sqrt{{\bar{n}}^2}},
\end{eqnarray}
the four vectors ${\bar{u}}^\mu$ and ${\bar{n}}^\mu$ used in the 
construction of the above tensors is defined as follows
\begin{eqnarray}
\bar{u}^\mu &=&\left(g^{\mu\nu}-\frac{Q^\mu Q^\nu}{Q^2}\right)u_\nu,\\
\bar{n}^\mu &=&\left(\tilde{g}^{\mu\nu}-\frac{\tilde{Q}^\mu\tilde{Q}^\nu}
{\tilde{Q}^2}\right)n_\nu,
\end{eqnarray}
where ${\tilde{g}}^{\mu\nu}=g^{\mu\nu}-u^\mu u^\nu$ and 
$\tilde{Q}^\mu=Q^\mu-(Q.u)u^\mu$. 
    
\subsection{Evaluation of the form factor $b(Q)$}
\label{b}
We will evaluate the real and imaginary
parts of the form factor 
$b$. Using Eq.\eqref{self_energy2}, the form factor b can be written upto $O[(q_fB)^2]$  as
\begin{eqnarray}
b(Q)=b_0(Q)+b_2(Q),
\label{formfactor_b}
\end{eqnarray}
where the form factors $b_0$ and $b_2$ are represented as follows: 
\begin{eqnarray}
b_0(Q)&=&\frac{u^\mu u^\nu}{\bar{u}^2}\Pi^{\mu\nu}_{(0,0)}(Q),
\label{formfactor_b0}\\
b_2(Q)&=&\frac{u^\mu u^\nu}{\bar{u}^2}[\Pi^{\mu\nu}_{(1,1)}(Q)+
2\Pi^{\mu\nu}_{(2,0)}(Q)].
\label{formfactor_b2}
\end{eqnarray}
\textbf{\underline{Form factor $b_0(Q)$ (order of $O[(q_fB)^0]$)}}:\\
\\
We will use the imaginary-time
formalism to calculate the form factor $b_0$, which 
is given by 
\begin{eqnarray}
b_0(Q)
&=&-N_f \frac{2g^2}{\bar{u}^2}\int\frac{d^3k}{(2\pi)^3}T\sum_n\frac
{\left[K^2+2k^2\right]}
{(K^2-m^2_f)(P^2-m_f^2)},\nonumber\\
&=&-N_f \frac{2g^2}{\bar{u}^2}[I_1(Q)+I_2(Q)],
\label{formfactor1_b0}
\end{eqnarray}
where we have neglected $m_f$ in numerator in the Hard 
Thermal Loop (HTL) approximation and $\int\frac{d^4K}{(2\pi)^4} 
\rightarrow iT\int\frac{d^3k}{(2\pi)^3}\sum_n$, 
after performing the frequency sum and retaining only the leading term in $T$, the $I_1$ and $I_2$ become
\begin{eqnarray}
I_1(Q)&=&\int\frac{d^3k}{(2\pi)^3}\frac{n_F(E_1)}{E_1},\\
I_2(Q)
&=&-\int\frac{d^3k}{(2\pi)^3}\left[\frac{n_F(E_1)}{E_1}
+q\cos\theta\frac{dn_F(E_1)}{dk}\frac{1}{i\omega-
q\cos\theta}\right].
\end{eqnarray}
where $n_F(E) = \frac{1}{\left[e^{\frac{(E - \mu)}{T}} + 1\right]}$ is the quark distribution function with chemical potential $\mu$. The evaluation of $I_1$ and $I_2$ leads to the real and imaginary parts of $b_0$ in the static limit
\begin{eqnarray}
{\rm \Re}~b_0(q_0=0)&=&N_f\frac{g^2T^2}{6}+N_f 2 g^2T^2 \hat{\mu}^2\\
\left[\frac{{\rm \Im}~b_0(q_0,q)}{q_0}\right]_
{q_0=0}&=&\left\lbrace\frac{g^2T^2N_f}{6}+ N_f 2g^2 T^2 \hat{\mu}^2\right\rbrace\frac{\pi}{2q}.
\end{eqnarray} 
We now add the gluon-loop contribution, which remains unaffected by the presence of the magnetic field, to the obtained quark-loop contributions. After adding both the contributions, the real and imaginary parts of $b_0$ are expressed as 
\begin{eqnarray}
{\rm \Re}~b_0(q_0=0)&=&\frac{g^2T^2}{3}\left\lbrace \left(N_c+\frac{N_f}{3}\right)+6N_f\hat{\mu}^2\right\rbrace,
\label{real_b0}\\
\left[\frac{{\rm \Im}~b_0(q_0,q)}{q_0}\right]_{q_0=0}
&=&\frac{g^2T^2}{3}\left\lbrace \left(N_c+\frac{N_f}{3}\right)+6N_f\hat{\mu}^2\right\rbrace\frac{\pi}{2q}.
\label{img_b0}
\end{eqnarray}
Thus, we observe that the form factor $b_0$ is independent of the magnetic field, as it appears at order $\mathcal{O}\!\left[(q_f B)^0\right]$, and depends only on the temperature of the medium and the chemical potential. In the absence of the magnetic field, the form factor $b_0$ coincides with the HTL longitudinal form factor $\Pi_L$ ~\cite{Weldon:PRD26'1982,Pisarski:PRL63'1989}.

\noindent\textbf{\underline{Form factor $b_2(Q)$ (order of $O[(q_fB)^2]$)}}:\\

\noindent Similar to the form factor $b_0$, here we will solve the 
form factor $b_2$, which is given by
\begin{eqnarray}
b_2(Q)&=&\sum_f \frac{i2g^2(q_fB)^2}{\bar{u}^2}\left[\int\frac
{d^4K}{(2\pi)
^4}\left\lbrace\frac{\left(2k_0^2-K_\parallel^2+m_f^2\right)}
{(K^2-m^2_f)^2(P^2-m_f^2)^2}
-\frac{\left(8k_0^2K_\perp^2\right)}{(K^2-m^2_f)^4(P^2-m_f^2)}\right
\rbrace\right],\nonumber\\
&=&-\sum_f \frac{2g^2(q_fB)^2}{\bar{u}^2}\int\frac{d^3k}{(2\pi)^3}
T\sum_n\left\lbrace\frac{K^2+k^2(1+\cos^2\theta)+m_f^2)}
{(K^2-m^2_f)^2(P^2-m_f^2)^2}\right.\nonumber\\&&\left.-
\frac{8(k^4+k^2K^2)(1-\cos^2\theta)}
{(K^2-m^2_f)^4(P^2-m_f^2)}\right\rbrace,
\end{eqnarray}
where we have used the spherical polar coordinate 
system for $k=(k\sin\theta\sin\phi,k\sin\theta
\cos\phi,k\cos\theta)$. In order
to solve the form factor $b_2$, we will use the method 
as shown in~\cite{karmakar:EPJC79'2019} , which gives 
\begin{eqnarray}
b_2(Q)&=&-\sum_f\frac{2g^2q_f^2B^2}{\bar{u}^2}\left[
\left\lbrace\frac{\partial}{\partial(m_f^2)}+\frac{5}{6}
m_f^2\frac
{\partial^2}{\partial^2(m_f^2)}\right\rbrace
\int\frac{d^3k}{(2\pi)^3}T\sum_n\frac{1}{(K^2-m_f^2)
(P^2-m_f^2)}\right.\nonumber\\&&\left.-\left\lbrace\frac
{\partial}{\partial(m_f^2)}+\frac{m_f^2}{2}\frac
{\partial^2}{\partial^2(m_f^2)}\right\rbrace
\int\frac{d^3k}{(2\pi)^3}T\sum_n\frac{\cos^2\theta}
{(K^2-m_f^2)(P^2-m_f^2)}
\right],
\end{eqnarray}
we will now perform the frequency sum to get the real and imaginary parts of $b_2$ in the static limit, which is obtained as~\cite{karmakar:EPJC79'2019}
\begin{eqnarray}
{\rm \Re}~b_2(q_0=0)&=&\sum_f\frac{g^2}{12\pi^2 T^2}(q_fB)^2
\sum_{l=1}^{\infty}(-1)^{l+1}l^2\cosh(2l\pi\hat{\mu})K_0(\frac{m_fl}{T})\\
\left[\frac{{\rm \Im}~b_2(q_0,q)}{q_0}\right]_
{q_0=0}&=&\frac{1}{q}\left[\sum_f\frac{g^2(q_fB)^2}{16\pi T^2}
\sum_{l=1}^\infty(-1)^{l+1}l^2K_0\left(\frac{m_f l}{T}\right)
\right. \nonumber\\ &&\left. 
-\sum_f\frac{g^2(q_fB)^2}{96\pi T^2}
\sum_{l=1}^\infty(-1)^{l+1}l^2K_2\left(\frac{m_f l}{T}\right)
\right. \nonumber\\ &&\left.
+\sum_f\frac{g^2(q_fB)^2}{768\pi}\frac{(8T-7\pi m_f)}{m_f^2 T}
\right].
\end{eqnarray}
where $K_0$ and $K_2$ are the modified Bessel functions of the second kind. After obtaining the real and imaginary parts of
the form factor $b_0$ and $b_2$, we can write the 
real and imaginary parts of the form factor $b$ using 
Eq.\eqref{formfactor_b} as follows
\begin{eqnarray}
{\rm \Re}~b(q_0=0)&=&\frac{g^2T^2}{3}\left\lbrace\left(N_c+\frac{N_f}{2}\right)+6N_f\hat{\mu}^2\right\rbrace
\nonumber\\&+&
\sum_f\frac{g^2}{12\pi^2 T^2}(q_fB)^2
\sum_{l=1}^{\infty}(-1)^{l+1}l^2\cosh(2l\pi\hat{\mu})K_0(\frac{m_fl}{T}),
\label{real_b}
\end{eqnarray}
\begin{eqnarray}
\left[\frac{{\rm \Im}~b(q_0,q)}{q_0}\right]_
{q_0=0}&=&\frac{g^2T^2}{3}\left\lbrace\left(N_c+\frac{N_f}{2}\right)+6N_f\hat{\mu}^2\right\rbrace\frac{\pi}{2q}
+\frac{1}{q}\left[\sum_f\frac{g^2(q_fB)^2}{16\pi T^2}
\sum_{l=1}^\infty(-1)^{l+1}l^2K_0\left(\frac{m_f l}{T}\right)
\right. \nonumber\\ &-&\left. 
\sum_f\frac{g^2(q_fB)^2}{96\pi T^2}
\sum_{l=1}^\infty(-1)^{l+1}l^2K_2\left(\frac{m_f l}{T}\right)
+\sum_f\frac{g^2(q_fB)^2}{768\pi}\frac{(8T-7\pi m_f)}{m_f^2 T}
\right],
~~\label{img_b}
\end{eqnarray}
where the real part of the form factor evaluated in Eq.\eqref{real_b} in the static limit is equivalent to the Debye screening mass $({\rm \Re}~b(q_0=0)=m_D^2)$ in the presence of a weak magnetic field with chemical potential, which has recently been calculated in~\cite{Bandyopadhya:PRD100'2019}.
\end{appendices}

\end{document}